\documentclass[doublecol]{epl2} 

\usepackage[bookmarks]{hyperref}
\usepackage{graphics}
\usepackage{exscale}
\usepackage{epsfig}
\usepackage[T1]{fontenc}
\usepackage{bm}
\usepackage{bbm}

\usepackage{version}

\usepackage{latexsym}

\renewcommand{\epsilon}{\varepsilon}

\newcommand{\integral}[3]{\!\int\limits_{#2}^{#3}\!\!{\rm d}#1\;}

\newcommand{\expval}[2]{ \langle  #1 #2\ \!\! \rangle}

\newcommand{\elcre}[2]{ c^{\dagger}_{#1,#2}}
\newcommand{\elann}[2]{ c_{#1,#2}}

\newcommand{\vk}{{\bm k}}

\newcommand{\Imag}{\mathrm{Im}}

\newcommand{\hc}{\mathrm{h.c.}}

\includeversion{commentst1} 

\begin{document}

\title{Competing interactions and symmetry breaking in the Hubbard-Holstein model} 
\author{Johannes Bauer}
\institute{Max-Planck Institute for Solid State Research, Heisenbergstr.1,
  70569 Stuttgart, Germany} 
\date{\today} 
\abstract{
Competing interactions are often responsible for intriguing phase diagrams in
correlated electron systems. Here we analyze the competition of instantaneous
short range Coulomb interaction $U$ with the retarded electron-electron
interaction induced by an electron-phonon coupling $g$ as described by the
Hubbard-Holstein model. The ground state 
phase diagram of this model in the limit of large dimensions at half
filling is established. The study is based on dynamical mean field  
theory combined with  the numerical renormalization group. Depending 
on $U$, $g$, and the phonon frequency $\omega_0$, the ground state
is antiferromagnetically (AFM) or charge 
ordered (CO).  We find quantum phase transitions from the AFM to CO state 
to occur when $U-\lambda\simeq 0$, where  $\lambda$ characterizes the phonon induced
effective attraction. The transition is continuous for small couplings and
large phonon frequencies $\omega_0$ and becomes discontinuous for large
couplings and small values of $\omega_0$.  
We comment on the possible relevance of this work for 
Ba${}_{1-x}$K${}_x$BiO${}_3$.
}


\maketitle

\section{Introduction}
A classical problem in condensed matter physics is
that of the competing effects of the Coulomb repulsion of the electrons with
the attraction generated by the electron-phonon coupling.
This appears most prominently in the theory of conventional
superconductivity. There, the relatively weak 
attraction mediated by the phonons wins against the Coulomb repulsion
with the help of retardation effects and the fact that the latter is
renormalized to a reduced value at the phonon scale \cite{MA62}.
Only with this pseudo potential effect phonon induced superconductivity is credible
in spite of the omnipresent Coulomb repulsion.
Beyond such weak coupling arguments the competition of instantaneous repulsion
and retarded attraction has only started to be explored in recent years with
the advent of reliable strong coupling methods.
In many materials  the low energy physics is dominated by different
competing - often strong - interactions, which need to be studied simultaneously. For 
example consider strongly correlated systems such as the high $T_c$ cuprates
\cite{Lea01}, fullerides \cite{gunnarsson}, manganites \cite{Mil98}, and
organic salts \cite{PM06} and their intriguing phase diagrams. 
The purpose of this letter is to analyze the competition of instantaneous
local Coulomb repulsion and the attraction mediated by phonons for different
coupling strengths, and establish the resulting ground states, allowing for
symmetry breaking in the magnetic and charge channel. 

As concrete model to study the competing effects we choose  the combination the fundamental
model of local electronic correlations, the Hubbard model \cite{Hub63}, with a classical
model of electron-phonon coupling, the Holstein model \cite{Hol59}.
Both models are probably oversimplified as to describe real materials in
detail, but they can serve to obtain insights into dominant interaction effects.
The combined Hubbard-Holstein (HH) Hamiltonian reads 
\begin{eqnarray}
  \label{hubholham}
  H&=&-t\sum_{i,j,{\sigma}}(\elcre i{\sigma}\elann
j{\sigma}+\hc)+U\sum_i \hat n_{i,\uparrow}\hat n_{i,\downarrow} \\
&&+\omega_0\sum_ib_i^{\dagger}b_i+g\sum_i(b_i+b_i^{\dagger})\Big(\sum_{\sigma}\hat
n_{i,\sigma}-1\Big). 
\nonumber
\end{eqnarray}
$\elcre i{\sigma}$ creates an electron at lattice site $i$ with spin $\sigma$,
and $b_i^{\dagger}$ a phonon with oscillator frequency $\omega_0$, $\hat
n_{i,\sigma}=\elcre i{\sigma}\elann i{\sigma}$. The electrons interact locally
with strength $U$, and their density couples to an optical phonon mode with
coupling constant $g$. Depending on $U$, $g$, $\omega_0$, and the filling
factor $n$, the model is expected to display a variety of different phases
including polaronic  normal (N) state behavior, antiferromagnetic (AFM),
charge (CO) and superconducting (SC) order. It is important to map out these
phases to see whether for a certain material an effective description in terms
of the fundamental HH Hamiltonian is sensible. We focus on the case $n=1$.

Due to the large number of coupled degrees of freedom no exact solution of
the HH model is available for the general case, and there are few analytical
methods which respect the quantum nature of the phonons and allow for
arbitrary coupling strengths $U, g$. 
For the pure electron-phonon problem perturbative schemes, such as
Migdal-Eliashberg theory, can be very successful; strong Coulomb interactions
can however not be treated.
For the combined HH model there has been a lot of progress in recent years in
one and infinite dimensions. In the $d=1$ situation
numerical methods can be applied with high accuracy and the phase
diagram could be established \cite{CH05,HC07,TTCC07,FHJ08}. To our knowledge for $d>1$ general phase
diagrams are still missing.  In the case of high dimensions the dynamical mean field theory
(DMFT)  \cite{GKKR96} becomes exact, and it can generate non-perturbative
solutions, such that it becomes the method of choice for our purpose of
studying arbitrary coupling strengths.

\begin{figure}[!t]
\centering
\includegraphics[width=0.4\textwidth]{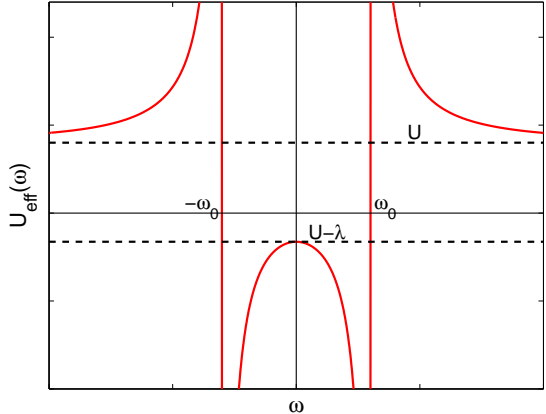}
\vspace*{-0.5cm}
\caption{(Color online) The dynamic effective interaction $U_{\rm
    eff}(\omega)$ as a function of $\omega$.}       
\label{Ueffom}
\end{figure}

The competition of the interactions can be seen in concise form when integrating out
the bosonic field, which yields an effective electronic interaction
\begin{equation}
  U_{\rm eff}(\omega)=U+\frac{2g^2\omega_0}{\omega^2-\omega_0^2}.
\label{ueffom}
\end{equation}
As depicted in Fig. \ref{Ueffom} 
for large $\omega$ the Coulomb repulsion $U$
is dominant, $\omega_0$ enters as a relevant energy scale at lower energy, and
for $|\omega| \le \omega_0$ the competition between the bare interactions
is most important. 

\noindent
The DMFT calculations deal with these competing interactions, and as a main
result of this paper the ground state phase diagram of the infinite
dimensional HH model at half filling emerges. This is likely to be good
approximation for the three dimensional case, where collective excitations are
not dominant.
\begin{figure}[!t]
\centering
\includegraphics[width=0.45\textwidth]{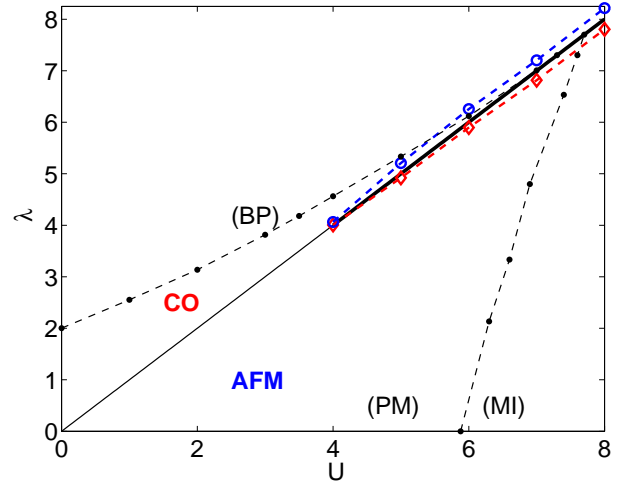}
\vspace*{-0.5cm}
\caption{(Color online) The phase diagram of AFM and
  CO in the $U$-$\lambda$-plane exemplary for $\omega_0=0.6 t$. The thin black
  line $\lambda \simeq U$ gives a continuous transition and the thick line a
  discontinuous one. A nonzero CO/AFM order parameter 
  was found above/below the  dashed line with diamonds/circles. The dashed
  lines with points gives the transition for phases with no long range order,
  a paramagnetic metallic (PM), bipolaronic (BP) and Mott insulator (MI).}       
\label{hfphasediag}
\end{figure}
\noindent
As shown in Fig. \ref{hfphasediag}, the transition from an AFM 
state to a CO state occurs near\footnote{For the given
  parameters, there is a small tendency towards $\lambda>U$ for the transition
  line, which can however hardly be resolved in the plot.}  $U_{\rm eff}= 0$, where $U_{\rm
  eff}=U-\lambda$ with $\lambda=2g^2/\omega_0$. For $ U_{\rm eff}<0$ we find a
CO and for $ U_{\rm   eff}>0$ an AFM
ground state. The phase boundaries for the phases without long range order,
the paramagnetic metal (PM), the Mott insulator (MI), and the bipolaron (BP)
insulator, are also shown in Fig. \ref{hfphasediag}. 
We see that only for large couplings the phase boundaries merge, but for smaller couplings other scales
are important as analyzed earlier for $\omega_0=0.2$ \cite{JPHLC04,KMOH04,KMH04}.
The deceptively simple result for the phase boundary of the ground state phase
diagram is expected in limiting cases, such as
the antiadiabatic one, $\omega_0\to \infty$ with $\lambda$ kept fixed,
where as seen from (\ref{ueffom}) the HH model reduces to the Hubbard model
with coupling constant $U_{\rm 
  eff}$. For the general case it is not a priori clear that 
this applies, as this depends on the mutual renormalization
effects of the couplings at low energy. However, our results indicate that
this remains 
valid for general $\omega_0$. In contrast to this universality the details and
order of the transition depend on both, the couplings and the value of
$\omega_0$. 

The infinite dimensional HH model has received considerable attention and has
been studied by DMFT. 
For instance, the formation of polarons \cite{KHE05} and the competition
between the Coulomb repulsion and the induced interaction \cite{SCCG05} were
studied in the symmetric phase.
Recently, the phase diagram of PM, BP and MI was established
\cite{JPHLC04,KMOH04}, where phases with long-range order were not allowed
for. Allowing for AFM order, the effect of the Coulomb repulsion on the
electron-phonon interaction was investigated \cite{SGKCC06}. Here we extend
these earlier calculations allowing for commensurate AFM {\em and} CO and
establish the full ground state phase diagram. In our calculations we also
studied SC solutions, but we found that for finite $\omega_0$ CO has lower
energy at half filling, thus SC does not appear in the phase diagram.

\section{Formalism}
For our calculations we assume a bipartite lattice with $A$ and $B$ sublattice, where
the matrix Green's function can be written in the form    
\begin{equation}
\underline{G}_{\vk,\sigma}(\omega) \!=\!
\frac1{\zeta_{A,\sigma}(\omega)\zeta_{B,\sigma}(\omega) -\epsilon_{\vk}^2}
\! \left(\!\!\!
\begin{array} {cc}
 \zeta_{B,\sigma}(\omega) & \epsilon_{\vk} \\
\epsilon_{\vk} & \zeta_{A,\sigma}(\omega)
\end{array}
\!\!\!\right),
\label{kgf}
\end{equation}
with $\zeta_{\alpha,\sigma}(\omega)=\omega+\mu_{\alpha,\sigma}-\Sigma_{\alpha,\sigma}(\omega)$,
$\alpha=A,B$, and $\vk$-independent self-energy \cite{MV89}.
For commensurate charge order, we have $\mu_{A,\sigma}=\mu-h_c$, $\mu_{B,\sigma}=\mu+h_c$ and
$\Sigma_{B,\sigma}(\omega)=Un-\Sigma_{A,\sigma}(-\omega)^*$, with
$n=(n_A+n_B)/2$, $n_{\alpha}=\sum_{\sigma}n_{\alpha,\sigma}$, where
$n_{\alpha,\sigma}=\expval{\hat n_{\alpha,\sigma}}{}$. For AFM order, one has
$\mu_{A,\sigma}=\mu-\sigma h_s$, $\mu_{B,\sigma}=\mu+\sigma 
h_s$, and the condition
$\Sigma_{B,\sigma}(\omega)=\Sigma_{A,-\sigma}(\omega)$.  We consider
spontaneously ordered solutions
where the symmetry breaking fields vanish, $h_c,h_s\to 0$.
The matrix elements of the local Green's function $\underline G(\omega)$ can
be calculated by integrating the matrix elements of (\ref{kgf}) over the
density of states, which we choose as semi-elliptic,
$\rho_0(\epsilon)=2\sqrt{D^2-\epsilon^2}/\pi D^2$.
We solve the effective impurity problem with the numerical renormalization
group \cite{Wil75,BCP08} (NRG) adapted to these cases with symmetry
breaking. In these calculations we have chosen the discretization parameter
$\Lambda=1.8$ and we keep around 1000 states at each iterations. The initial
bosonic Hilbert space is restricted to a maximum of 50 states. This is 
justified by the phonon occupation $n_{\rm ph}$ which does not exceed values of 
$n_{\rm ph}\simeq 10$ except when $\lambda$ is much larger than $U$.

In the AFM case the $A$-sublattice magnetization,
$\Phi_{\rm afm}=m_A=(n_{A,\uparrow}-n_{A,\downarrow})/2$ serves as an order parameter.
For CO we define  $\Phi_{\rm co}=(n_A-1)/2$.
To identify the ground state of the system, we must compute
the total ground state energy per lattice site, $E_{\rm   tot}=\expval{H}{}/N$, of the
HH Hamiltonian (\ref{hubholham}) in the different phases.
 This gives generally, 
\begin{equation}
  E_{\rm tot}=E_{\rm kin}+E_U+E_{\rm ph}+E_g. 
\label{energies}
\end{equation}
The first term is the kinetic energy.
which reads 
\begin{equation}
  E_{\rm kin}=
\sum_{\sigma}\integral{\epsilon_{\vk}}{}{}\rho_0(\epsilon_{\vk})\epsilon_{\vk}
\integral{\omega}{}{}f(\omega)\rho_{AB,\vk,\sigma}(\omega),  
\end{equation}
where $\rho_{AB,\vk,\sigma}(\omega)=-\Imag G_{AB,\vk,\sigma}(\omega)/\pi$ for
the offdiagonal Green's function in (\ref{kgf}) and
$f(\omega)$ is the Fermi function.
In the non-interacting case it can be evaluated analytically 
with 
$\rho_{\vk,\sigma}(\omega)=\delta(\omega-\epsilon_{\vk}+\mu)$,
and we find for half filling, $\mu=0$, $E_{\rm kin}^0=-4D/3\pi$, which for
$D=2$ is $E_{\rm kin}^0\simeq-0.8488$. This can be used as reference energy.
The interaction energies $E_U$, $E_g$ can be calculated from expectation values.
We have
\begin{equation}
  E_U=\frac{U}2\sum_{\alpha}\expval{\hat n_{\alpha,\uparrow}\hat
    n_{\alpha,\downarrow}}{},\;
  E_g=\frac g2\sum_{\alpha}
  \expval{(b_{\alpha}+b_{\alpha}^{\dagger})(\hat n_{\alpha}-1)}{}.
\nonumber
\end{equation}
We distinguish between $A$- and $B$-sublattice values, which are equal in the
AFM case, but not for the CO case.
The third term for the total energy is generally given by the
expectation value of the number of excited phonons, $E_{\rm
  ph}=\omega_0\expval{b^{\dagger}b}{}$.  The detailed behavior of the various
contributions to the energy in the different phases as well as static and
dynamic response functions will be discussed elsewhere \cite{BH10pre}.

\section{Results}
\begin{figure}[!t]
\centering
\includegraphics[width=0.45\textwidth]{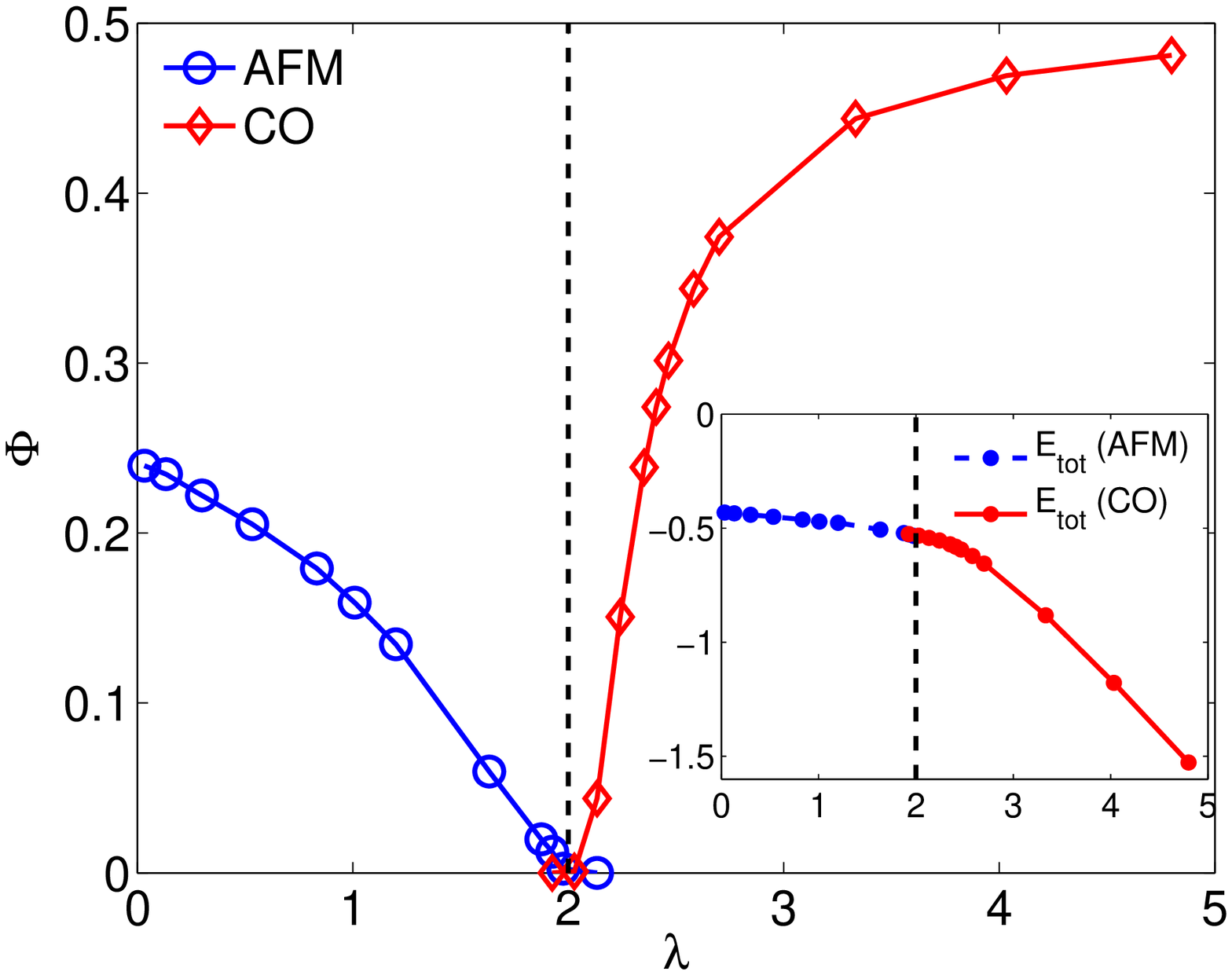}
\includegraphics[width=0.45\textwidth]{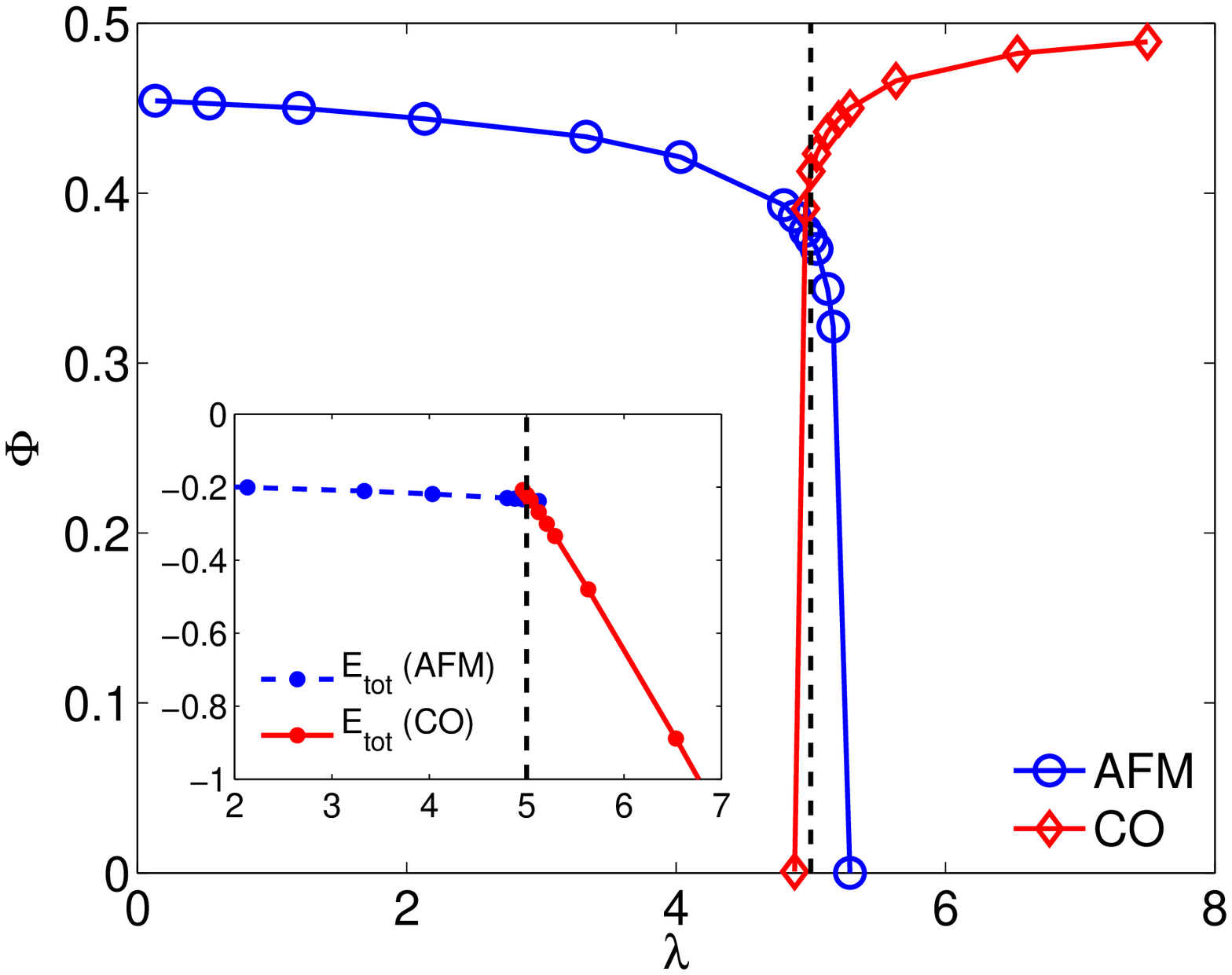}
\vspace*{-0.5cm}
\caption{(Color online) The expectation values $\Phi_{\rm afm}$ and $\Phi_{\rm co}$ for $U=2$ (top) and
  $U=5$ (bottom) as a function of $\lambda$. The total energy can be seen in
  the inset.}       
\label{phi_lamdepU25}
\end{figure}

The half bandwidth $D=2t=W/2$ is chosen as 2 in the following to set the energy scale. 
In Fig. \ref{hfphasediag}  the phase diagram is
shown for $\omega_0=0.6$ to exemplify the behavior neither too close to the
adiabatic nor to the antiadiabatic regime. We carried out numerous
calculations for other values of $\omega_0$ and  apart from the location of the
point separating continuous and discontinuous transitions similar behavior was
found for other choices. Notice that for physical optical phonons ($\omega_0\sim0.01W-0.2W$) this
value is rather large and the later presented results for $\omega_0=0.2$ can serve
as a better guideline.
Limiting cases of the phase   
diagram are known and easily understood on a qualitative level. Along the
$U$-axis, the pure repulsive 
Hubbard model (at half filling on a bipartite lattice) is known to be
AFM ordered at weak coupling \cite{Don91}, and this order is smoothly connected
to the strong coupling Heisenberg AFM \cite{ZPB02,BH07c}.
Along the $\lambda$-axis, the pure Holstein model has a charge ordered ground state for
$g>0$ and $\omega_0>0$ \cite{FJS93}, where the limits of weak and strong
coupling are smoothly connected. 
For finite $U$ and $g$ we find that the transition line is approximately
given by the line $\lambda\simeq U$.  
For the cases of weak and intermediate coupling ($U<W$) the order parameters become very
small in our calculation close to the line $U_{\rm eff}= 0$. An example is
shown for fixed $U=2$ in 
Fig. \ref{phi_lamdepU25} (top), where the order parameters are plotted as a
function of $\lambda$. 

\noindent

\begin{figure}[!htbp]
\centering
\includegraphics[width=0.45\textwidth]{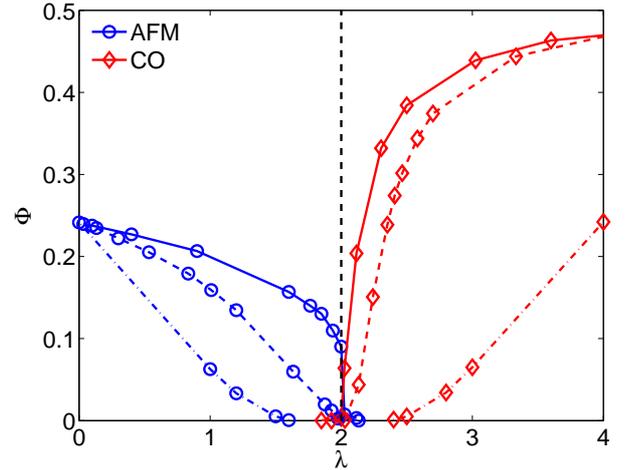}
\caption{(Color online) The expectation values $\Phi$ for fixed $U=2$ varying
  $\lambda$. We show the results for  $\omega_0=0.2$ (full line), 
  $\omega_0=0.6$ (dashed line), and $\omega_0\to\infty$ (dot-dashed line).}       
\label{phi_lamdepU2varwo1}
\end{figure}

Near $\lambda=U$ the ordering scale is very small
($<10^{-3}$) and can not be resolved well in our DMFT-NRG
calculations, which then do not converge sufficiently well. The
transition appears to be continuous, as all 
relevant response quantities change continuously. There are strong indications that
it occurs directly from an ordered to an ordered state and no intermediate
regime exists. We draw this conclusion by considering the behavior for
different values of $\omega_0$ (see also Fig. \ref{phi_lamdepU2varwo1}) .
For $n=1$ in the antiadiabatic case, $\omega_0\to\infty$, it is known that for any
finite $U_{\rm eff}>0$ the system is AFM ordered \cite{Don91} and for $U_{\rm
  eff}<0$ in the CO or SC state \cite{MRR90}. We have therefore a continuous
transition from an ordered to an ordered state at 
$U_{\rm eff}=0$ on varying $U$ or $\lambda$. In the DMFT-NRG calculations we
find that the smaller $\omega_0$ is the larger the order parameters near the
transition become (see Fig. \ref{phi_lamdepU2varwo1}). Thus we conclude that this 
transition scenario persists for weak coupling and finite $\omega_0$.
Mean field calculations in the adiabatic limit, $\omega_0\to 0$, also support the picture of a
direct transition between ordered states, however, the transition is
discontinuous then.
From NRG calculations in the normal state we can
calculate a local effective quasiparticle interaction $U^r$ by comparing the
magnitude of the lowest
two-particle excitation energy with twice the one-particle excitations energy at
the fixed point \cite{BH07c}. $U^r$ changes sign
approximately at $U_{\rm eff} = 0$, which is consistent with a change of ground states
there. To examine in detail what happens exactly at the 
transition at weak coupling and to analyze the critical properties 
requires an effective theory specifically for this transition.

For larger couplings we have included in Fig. \ref{hfphasediag} a dashed line
($\diamond$) above which DMFT-NRG finds solutions with finite $\Phi_{\rm co}$
and another one ($\circ$) below which $\Phi_{\rm afm}$  is well finite. 
An example 
for this behavior is shown in Fig. \ref{phi_lamdepU25} (bottom) for $U=5$,
where a much sharper behavior is seen.
The calculation of the total ground state energy (see inset of
Fig. \ref{phi_lamdepU25}) shows that also 
here the transition occurs approximately at $U_{\rm eff}\simeq 0$. 
A number of quantities such as the double occupancy
$\expval{\hat n_{\uparrow}\hat n_{\downarrow}}{}$ show discontinuities at the transition. The
total energy is a continuous function of $\lambda$, but it displays a kink at
the transition, such that a first derivatives will be discontinuous. 
The transition can be studied as a function of $U$ for
fixed $\lambda$, and a very similar picture emerges for weak and strong
coupling. For details we refer to \cite{BH10pre}.

As already alluded to, apart from the coupling strength the behavior near the
transition also depends on the phonon 
frequency. In the adiabatic limit, $\omega_0\to 0$, the phonons are
represented by a static field, and one does not find a substantial modification
of the ordered state as long as $U_{\rm eff}$ does not change sign. When this happens the order
changes abruptly and we find a discontinuous transition.  
\noindent
In the antiadiabatic limit we deal with an effective Hubbard model with $U_{\rm eff}$. The bare coupling
becomes very small close to the transition, where it changes sign. This gives
then behavior of a renormalized mean field theory with an exponential
behavior of the order parameter, which goes to zero for $U_{\rm eff}\to 0$,
and thus a continuous transition. Our calculations access the behavior between these limiting
cases and we give examples for the behavior of $\Phi$ for fixed $U$ and
varying $\lambda$ for  $\omega_0=0.2,0.6$ and $\omega_0\to\infty$
in Fig. \ref{phi_lamdepU2varwo1}. We can see that the order for finite $\omega_0$ is
always larger than in the $\omega_0\to\infty$ case. Our results give
discontinuous transitions upon varying $\lambda$ for $U\ge 4$ for
$\omega_0=0.6$, whilst for $\omega_0=0.2$ already for $U\ge 3$.

\begin{figure}[!htbp]
\centering
\includegraphics[width=0.23\textwidth]{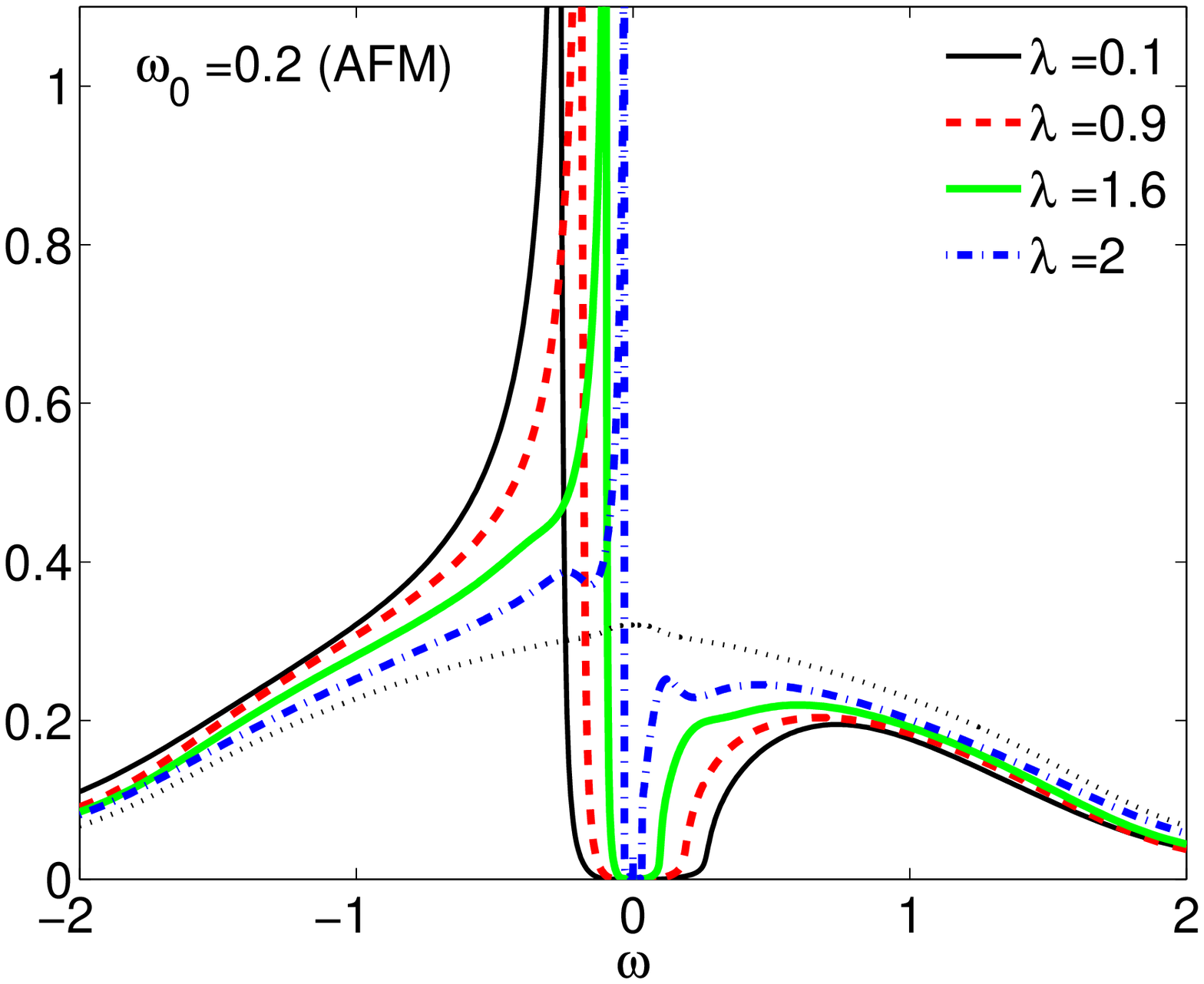}
\includegraphics[width=0.23\textwidth]{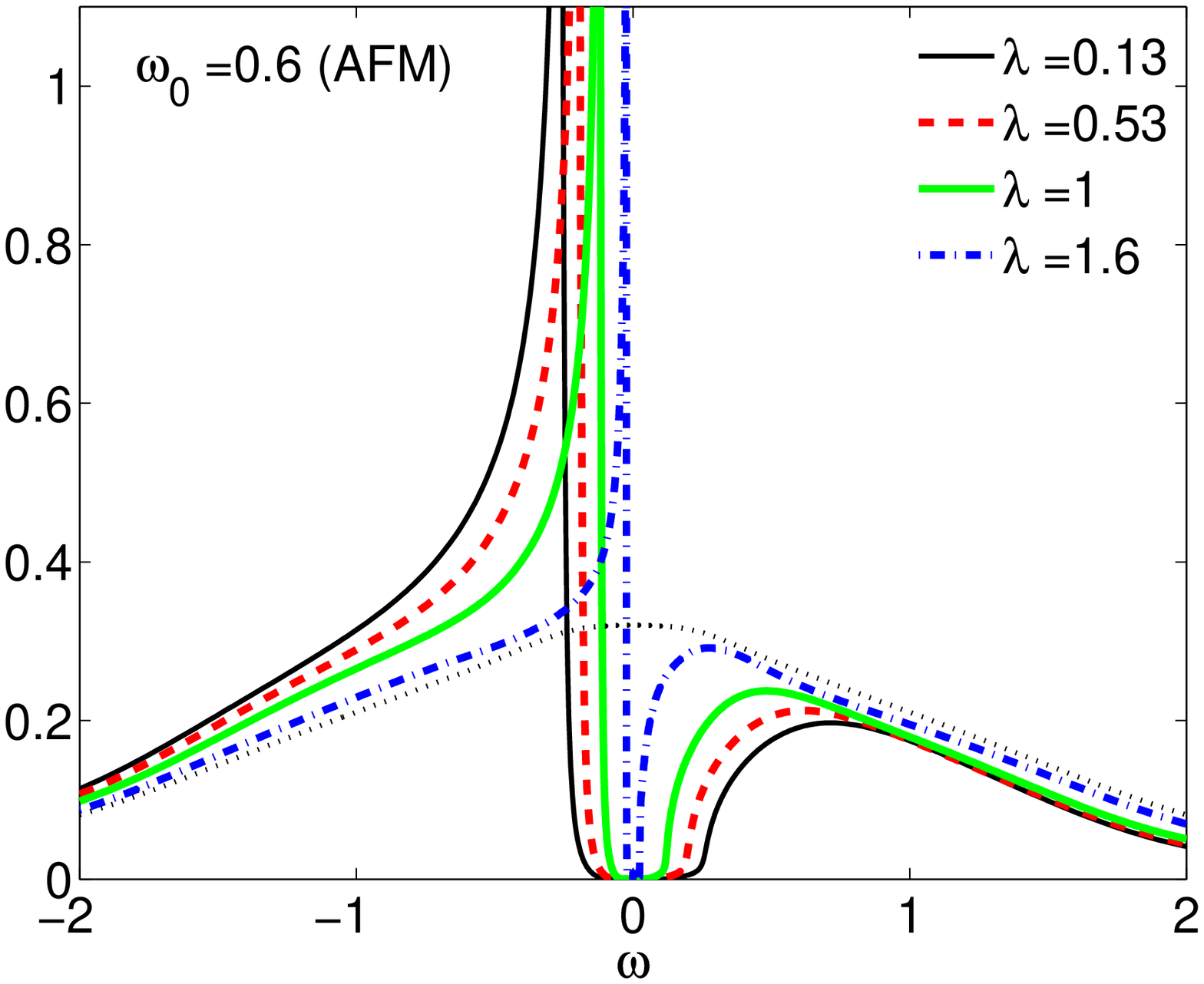}
\includegraphics[width=0.23\textwidth]{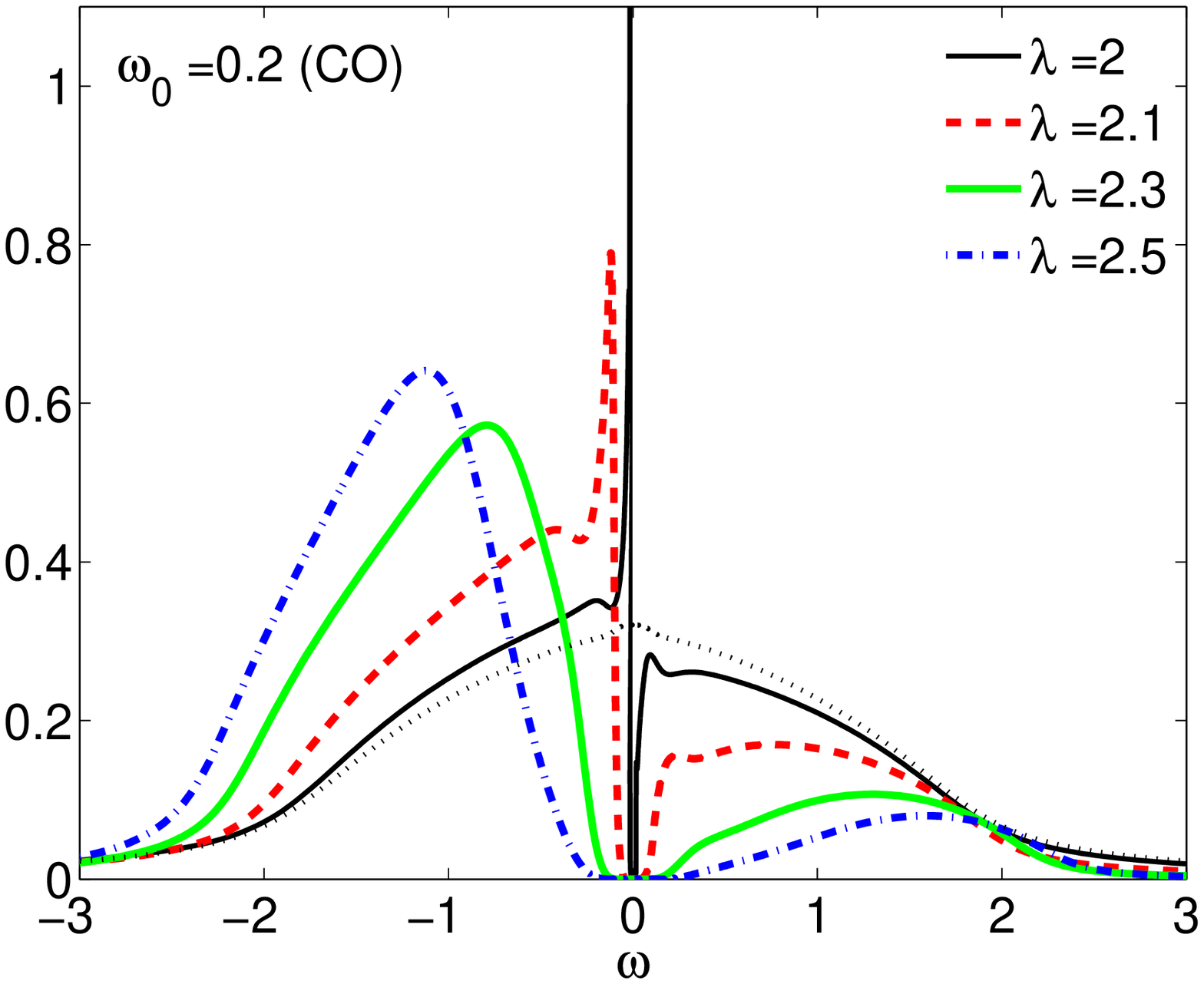}
\includegraphics[width=0.23\textwidth]{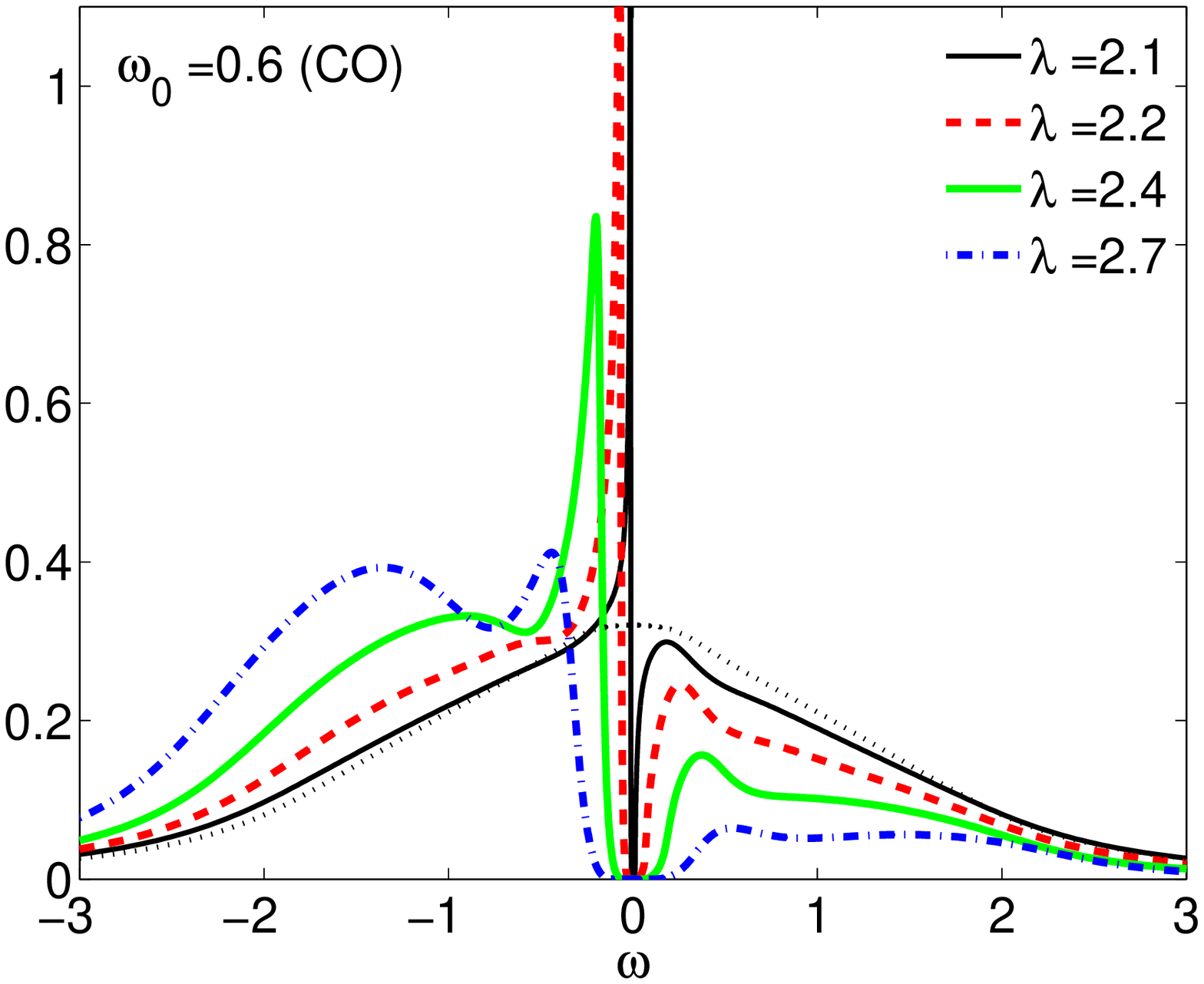}
\caption{(Color online) The local $A$-lattice (majority) spectral functions
  $\rho_{A,\uparrow}(\omega)$ in comparison AFM (upper part) and CO state
  (lower part) for $\omega_0=0.2,0.6$ for fixed $U=2$ and various $\lambda$
  approaching the transition. The dotted line gives the spectral function
  in the normal state without symmetry breaking for $U=\lambda$ for comparison.}       
\label{elspec_varlambdaU2difwo}
\end{figure}

More insight into the properties of the system are provided by the electronic
spectral function. Here we consider the (majority) local lattice Green's function
$G_{A,\uparrow}(\omega)$, which is given by the momentum sum of the diagonal
element of (\ref{kgf}), and its spectral function
$\rho_{A,\uparrow}(\omega)=-\Imag
G_{A,\uparrow}(\omega)/\pi$. $\rho_{\alpha,\sigma}(\omega)$ is symmetric at
half filling in the normal state but becomes asymmetric in
situations with symmetry breaking.  Note that at half filling the spectra for
minority spin in the AFM case, $\rho_{A,\downarrow}(\omega)$, and for the
$B$-lattice for the charge order, $\rho_{B,\uparrow}(\omega)$, can be obtained
from $\omega\to -\omega$. 
In Fig. \ref{elspec_varlambdaU2difwo} we show the spectral function for $U=2$
and various values of $\lambda$ approaching the transition in the AFM and CO
state for $\omega_0=0.2$ and $\omega_0=0.6$. For comparison we have added the normal state
spectral function for $U=\lambda$ as a dotted line in all cases.

\noindent

In the AFM state for small $\lambda$, the spectra fit well to the mean field
description, where a square root divergence is found below the gap and square
root increase above the gap.\cite{ZPB02} The higher energy parts are little
modified for the case of $U=2$ apart from the broadening of the band edges,
but no features which can be attributed to the phonons can be identified.  
When interpreting the spectra one has to take into
account the broadening and the limited energy resolution of the NRG at higher
energies, which limit the accuracy. 
On increasing $\lambda$, the AFM order and magnitude of the spectral gap decreases. The electron
phonon coupling is effective here in screening the repulsive $U$-term. No polaronic
features can be identified in the spectra as the coupling is fairly weak.
For $\omega_0=0.6$, the general form of the spectrum remains the same on
increasing $\lambda$, whereas
for $\omega_0=0.2$ small features near the gap emerge. In this case one can see that the
ordered solution and spectral gap persist for values of $\lambda$ up to $\sim
U$, whilst for $\omega_0=0.6$ the gap  vanishes earlier on increasing
$\lambda$ (see also Fig. \ref{phi_lamdepU2varwo1} for $\Phi$).  

In the CO state close to the transition, $\lambda\ge U$, similar spectral
characteristics of a weak coupling instability at the Fermi surface
($\omega=0$) as in the AFM case can be seen both for $\omega_0=0.2$ and
$\omega_0=0.6$. In the limit $\omega_0\to\infty$ we would find 
very similar behavior in the CO state on increasing $\lambda$ as in the AFM
state on increasing $U$, but as illustrated around Eq. (\ref{ueffom}) the
phonon-mediated attraction has an $\omega$-dependent dynamic form, which leads
to different behavior. One can identify a pronounced quasiparticle peak for interaction
values near the transition which however
becomes suppressed for larger values of $\lambda$. This suppression can be
partly due to the broadening in the NRG procedure as discussed in detail for
superconducting solutions \cite {BHD09}.
On increasing $\lambda$  a principal peak develops in the CO spectrum. Its position,
is fairly well described by the spectral shift of fully polarized mean field
theory, $\Delta_{\rm mf}=U n^{A}_{-\sigma}-2\lambda \Phi_{\rm co}$, for
instance, $\Delta_{\rm mf}=1.5$ for $\lambda=2.5$ and $U=2$ (cf. CO spectrum
for $\omega_0=0.2$). 

We make a few general remarks concerning the phase diagram which has been established.
It is remarkable that the phase diagram of the HH model in one dimension
\cite{HC07} is similar to Fig. \ref{hfphasediag}. There one finds a Mott
insulator with strong antiferromagnetic correlations, but {\em no} long range
order,  when $U_{\rm  eff}>0$, and a Peierls, charge ordered, insulator for $U_{\rm eff}<0$. There
is, however, a metallic region with finite spin gap, but no charge gap in
between these two phases \cite{CH05,HC07,FHJ08}. For larger $U$ this
intermediate region shrinks until one finds a direct first order Mott-Peierls
transition, similar to the present observation. A major difference with the
high dimensional results is the real symmetry breaking in our case as well as
the existence of the intermediate region, for which we find no indication here.

We comment on the possible relevance of this work for the three dimensional
compound Ba${}_{1-x}$K${}_x$BiO${}_3$ bearing in mind that there are still
controversial issues. According to band theory the compound would
be metallic for $x=0$ with a half filled Bi 6$s$ band, but it shows CO
\cite{HDRJPZ89} accompanied by a  
lattice distortion, which hints towards a strong coupling of
the electrons to the lattice. For $x>0.35$, SC with fairly high $T_c\approx 30$
K appears \cite{HDRJPZ89}. A
strong coupling to an oxygen breathing mode at roughly $70$meV is thought to
play an important role \cite{Lea89,CYKSNU01}. Thus, it has been
proposed that the Holstein or HH model with effective bandwidth $W\approx 4$eV
can provide an approximate description of the compound and account for the prominent features of the phase
diagram including CO and SC \cite{VNW92,Fre94,BVL95}. Estimates for the
dimensionless electron-phonon coupling constant, denoted by $\bar\lambda$ here,
based on density functional calculations  vary substantially, $\bar\lambda=0.4-1.5$ 
\cite{SSM90,LMRJAM91,MS98b,CYKSNU01}. If we
take $\bar\lambda=\lambda\rho_0(0)\simeq 1$, such that  $\lambda\simeq 3.14 t$,
then according to the phase diagram in Fig. \ref{hfphasediag} the
experimentally observed CO state can be explained only if the residual local
repulsion $U$ is significantly less than $\lambda=0.79 W$. This is consistent
with the estimates for a residual positive $U$ in an effective 
one-band model $U/W\approx 0.2$ \cite{VW96}. With the latter estimate for $U$ even for
$\bar\lambda=0.4$ \cite{MS98b} a CO state is stable for $x=0$. 
However, for such
small $\bar \lambda$ it is difficult to explain the high $T_c$ for
superconductivity in the conventional theory. These issues, including
appropriate values for $\bar\lambda$ for different fillings, still have
to be clarified by future research. Once suitable values are found, the approach
presented here or refinements of it can make predictions for quantities such as the spectral gap or
the transition temperature for the CO and SC state.

A question raised in this work is why the transition
between the two ordered states occurs for $U\simeq\lambda$ in such a
generality. Both bare interactions produce opposite effects for the
interactions of low energy quasiparticles. The fact that the sign
of the bare values $U-\lambda$ is reproduced for the effective quasiparticle
interaction $U^r$ suggests that $U$ and $g^2$ are renormalized in the
same way. A reason for this is the specific form of the electron-phonon
coupling to the local density and the phase space restriction at half
filling. This is different away from half filling and for couplings to other modes such 
as of the Jahn-Teller type \cite{gunnarsson}. 
A renormalization group study \cite{TTCC07}
has examined the competing interactions for the HH model in 
one dimension. Extensions to higher dimensions would be of great interest.

\section{Conclusions}
We analyzed competing interaction effects and established the ground state
phase diagram of the infinite dimensional Hubbard-Holstein 
model at half filling with a transition line $U\simeq\lambda$. This implies
that unlike superconductivity, where retardation is effective, phonon induced
CO prevails only if $\lambda$ exceeds the Coulomb repulsion.
However as seen in 
Fig. \ref{phi_lamdepU25}, a small excess can be sufficient to have a strongly
ordered state. We also identified a point separating  
continuous and discontinuous transitions. A continuous quantum phase transition between 
different ordered state is rare and could be of interest for further study.
We expect the presented phase diagram to be general as it holds for finite
$\omega_0$ as well as in the adiabatic and antiadiabatic limiting cases. From
the similarities with the one-dimensional phase diagram, and recent results
in $d=2$ \cite{KB08} (adiabatic limit), we conjecture that the form of the
phase diagram  possesses validity also in two and three dimensions. Extensions to
finite temperature and other fillings would be desirable, where the
competition can be studied in the case of superconductivity.

\noindent
I wish to thank O. Gunnarsson, A.C. Hewson, G. Sangiovanni, and R. Zeyher for
helpful discussions, and W. Koller and D. Meyer for their earlier
contributions to the development of the NRG programs.

\bibliographystyle{eplbib}
\bibliography{artikel,biblio1}

\end{document}